\documentclass[a4paper,11pt]{article}
\pdfoutput=1 

\usepackage{jheppub} 

\usepackage[T1]{fontenc} 

\usepackage{setspace}

\def\be{\begin{equation}}
\def\ee{\end{equation}}

\def\bea{\begin{eqnarray}}
\def\eea{\end{eqnarray}}

\def\mes[#1]{d^{3}{#1}}

\def\Tr{\text{Tr}}

\title{On The Entanglement Entropy For Gauge Theories}

\author[a]{Sudip Ghosh,}
\author[b]{Ronak M Soni,}
\author[b]{ and Sandip P. Trivedi}

\vspace{5mm}

\affiliation[a] {\it International Centre for Theoretical Sciences, Tata Institute of Fundamental Research,\\ IISc Campus, Bangalore, 560012, India }
\affiliation[b]
{\it Department of Theoretical Physics,
 Tata Institute of Fundamental Research,\\  Colaba, Mumbai, 400005, India}

\vspace{1cm}

\emailAdd{sudip112phys@gmail.com}
\emailAdd{ronak@theory.tifr.res.in}
\emailAdd{trivedi.sp@gmail.com}

\abstract{We propose a definition for the entanglement entropy of  a gauge theory  on a spatial lattice.  Our definition  applies to 
any subset  of links in the lattice, and is valid for both Abelian and Non-Abelian gauge theories. For  $\mathbb{Z}_N$ and $U(1)$ theories, without matter, our definition agrees with a particular case of the definition given by Casini, Huerta and Rosabal. We also argue that  in general, both for  Abelian and Non-Abelian theories, our definition agrees with the entanglement entropy calculated using a definition of the replica trick. Our definition, however, does not agree with some  standard ways to measure  entanglement, like the number of Bell pairs which can be produced by entanglement distillation. }

\preprint{\parbox{3cm}{TIFR/TH/15-03
 }}

 \arxivnumber{1501.02593}

\begin{document} 
\maketitle
\flushbottom

\section{Introduction}
\label{intro}
Entanglement entropy is an important measure  of quantum correlations in a system and provides  a way to quantify  the ``essential weirdness''  which arises in quantum mechanics. 
Gauge theories are a central paradigm in modern physics. Three of the four fundamental forces in nature are described by gauge theories. And via holography, they  are  important now  in our study of the fourth fundamental force, gravity.  Also, gauge theories   are  an increasingly  important corner stone  of condensed matter physics.  It is therefore important to try and provide a precise definition of entanglement entropy for gauge theories. 

This is however not straightforward. 
In a theory, like a spin system or a scalar field theory, which has  localised physical excitations,  the Hilbert space of states admits a tensor product decomposition in terms of local degrees of freedom. To obtain the entanglement entropy associated with a region of space we first  trace over the degrees of freedom outside to  obtain a density matrix. And then calculate the   von Neumann entropy of this density matrix to get  the entanglement entropy. 
In a gauge theory, unlike the  spin system or  the  scalar field theory, however, the physical degrees of freedom are not always local in space-time. 
For example, in a pure gauge theory without matter, they correspond to closed loops of electric or magnetic flux, which are non-local. 
As a result, the Hilbert space of physical gauge-invariant states does not admit a local tensor product description, and without such a decomposition it is not clear how to define a density matrix 
associated with some region of space and 
obtain the entanglement entropy. 

In this note we attempt to give such a definition for gauge theories.  Since the entanglement entropy is known to suffer from  short-distance divergences in field theories, we work with gauge theories living on a spatial lattice.
Motivated by the work of Casini, Huerta and Rosabal, \cite{CHR}, (CHR), see also \cite{Rad}, we then ask a slightly different question from the one posed above. Instead of a spatial region we consider a subset of links in this spatial lattice and 
ask about the entanglement entropy of these links. 
We find that this question lends itself to a precise  and elegant answer. 

The central point leading to our definition is as follows. While the Hilbert space of gauge-invariant states does not admit a tensor product decomposition in terms of local degrees of freedom, a larger Hilbert space ${\cal H}$ can be  defined admitting such a decomposition. This larger Hilbert space is obtained by taking the tensor product of the Hilbert spaces on which the individual link excitations live, and therefore arises quite naturally in the gauge theory. It  is larger because it  includes both gauge-invariant and non-gauge-invariant states. By regarding any gauge-invariant pure state $|\psi\rangle\in{\cal H}$ we can then obtain a density matrix, $\rho_{in}$, for the links of interest, which we call the ``inside'' links, 
  by tracing over the outside ones. Our  definition of the entanglement entropy is then simply given by  taking the von Neumann entropy of this density matrix: 
\be
\label{eein}
S_{EE}=-\Tr_{{\cal H}_{in}}\rho_{in} \log \rho_{in},
\ee
where the trace is taken over the Hilbert space, ${\cal H}_{in}$,  which is the tensor product of the Hilbert spaces  of the inside links. 

The definition above applies to  both Abelian and Non-Abelian gauge theories. Also, while for simplicity we will focus on $2$ spatial dimensions in   this paper,  the definition  holds in higher spatial dimensions as well. 

 We find  that in the  Abelian $\mathbb{Z}_N$ and $U(1)$ gauge theories, without matter, our definition agrees with the electric centre choice  for the entanglement entropy given in \cite{CHR}. 
 For  the Non-Abelian case  we find things are similar, and our definition can also be related to an electric centre type prescription, although the different sectors do not correspond to different values of electric flux. 
 
 Importantly, we also find  that for both the Abelian and Non-Abelian cases our definition agrees with the result for the entanglement entropy obtained by using a definition of he replica trick. 

Our definition has one unsatisfactory feature, though. 
In  quantum information theory  the entanglement of a bipartite system, i.e. a system with two parts,  is often quantified by comparing it to the entanglement of a standard system, e.g.,  a collection of Bell pairs. 
The comparison can be done, by converting the entanglement between the two parts  into  the entanglement of  a set of Bell pairs, as happens in entanglement distillation.
Or by consuming the entanglement in a set of Bell pairs to quantum teleport one of the two parts,  as happens in entanglement dilution. It turns out,  that the entanglement entropy we have defined does  not equal the number of Bell pairs which can be produced or consumed in this way. This fact was noticed by CHR already for the Abelian case, and it  continues 
to be also  true, with an interesting new aspect,  in the Non-Abelian case. 

The underlying reason for this difference is the   presence of non-local gauge-invariant operators, and excitations associated with these operators in gauge theories.  The protocols for entanglement distillation or dilution involve only local operations (and classical communication)
and this turns out to be  not enough for  converting  the full entropy  into  Bell pairs.  Since the presence of non-local gauge-invariant operators is a central feature in gauge theories, as was emphasised at the outset above, we suspect that such a limitation might be more generally true in other attempts to define the entanglement entropy in gauge theories as well.


The paper is structured as follows. We start first by discussing the $\mathbb{Z}_2$ gauge theory without matter in $2$ spatial dimensions  in section \ref{z2}.  Some general properties for  our definition of the entanglement entropy are discussed in the context of the $\mathbb{Z}_2$ example in section \ref{properties}. Then we consider the $U(1)$ case in section \ref{ua}. This sets the stage for the Non-Abelian case, which we discuss specifically for the $SU(2)$ example in section \ref{su2}. 
Connections to the replica trick are analysed in section \ref{replica}, followed by a discussion of additional aspects including the relation to more operational definitions of entanglement, in section \ref{more}.   We end in \ref{conc} with some conclusions and open directions.

Before ending the introduction let us also mention some of the  other relevant literature. Background material on entanglement in field theories that we found useful is in \cite{Srednicki}, \cite{HLW},
\cite{DFM}, 
\cite{CC1}, \cite{CC2}, \cite{CH1}, and \cite{S1}. 
For issues related to entanglement in gauge theories some references include \cite{Kabat}, \cite{HIZ}, 
\cite{DD1}, \cite{BP},  \cite{D1}, \cite{D2}, \cite{S2},  \cite{D3}.

\section{ The $\mathbb{Z}_2$ Lattice  Gauge Theory }
\label{z2}
Our starting point is a $\mathbb{Z}_2$ lattice gauge theory  without matter,  one of the simplest examples of a  gauge theory. For definiteness we work in $2+1$ dimensions.
We will  consider the theory in the Hamiltonian formulation, where time is continuous, and the two spatial directions are on a lattice, see \cite{Kogut}, which for simplicity we take to be square.

A state in this theory is defined, at an instant of time, on the two-dimensional spatial lattice. There is a two dimensional Hilbert space, corresponding to one qubit,  associated with each link of this lattice. 
A basis for this Hilbert space can be taken to be  $|\pm 1\rangle$,
with
\be
\label{defbasis}
\sigma_3 |\pm1\rangle=\pm 1|\pm 1\rangle
\ee
where $\sigma_3$ is the standard Pauli operator.   This basis is orthonormal, meeting conditions $\langle+1|+1\rangle=1, \langle+1|-1\rangle=0$, etc. Gauge transformations act on vertices of the lattice. A gauge transformation ${\cal G}_{V_i}$ acting on vertex $V_i$ flips the values of all link variables emanating from $V_i$. In the basis introduced above we have 
\be
\label{repgi}
{\cal G}_{V_i}= \prod \sigma_1
\ee
where the product is taken over all links which emanate from $V_i$. 
Gauge-invariant states are invariant under all gauge transformations, i.e., if $|\psi\rangle$ is a gauge-invariant state, then, 
\be
\label{gistrans}
{\cal G}_{V_i} |\psi\rangle =|\psi\rangle \
\ee
for all gauge transformations
${\cal G}_{V_i}$.
Similarly gauge-invariant operators are invariant  under gauge transformations.
If ${\hat O}$ is a gauge-invariant operator, 
\be
\label{gino}
{\cal G}_{V_i} {\hat O} {\cal G}_{V_i}^{-1}={\hat O}.
\ee

Gauge-invariant operators in the theory include Wilson loops which are the products of $\sigma_3$  operators for links forming closed plaquettes, and the $\sigma_1$ operators on each link. 
These measure the discrete analogue of  the magnetic  and electric fluxes respectively  in this theory. The most general gauge-invariant operator can be constructed by taking products of these operators. In the discussion below we refer to the $\sigma_1$ operators, somewhat loosely, as the electric flux operators.

\begin{figure}[t]
  \begin{center}
    \includegraphics[height=60 mm]{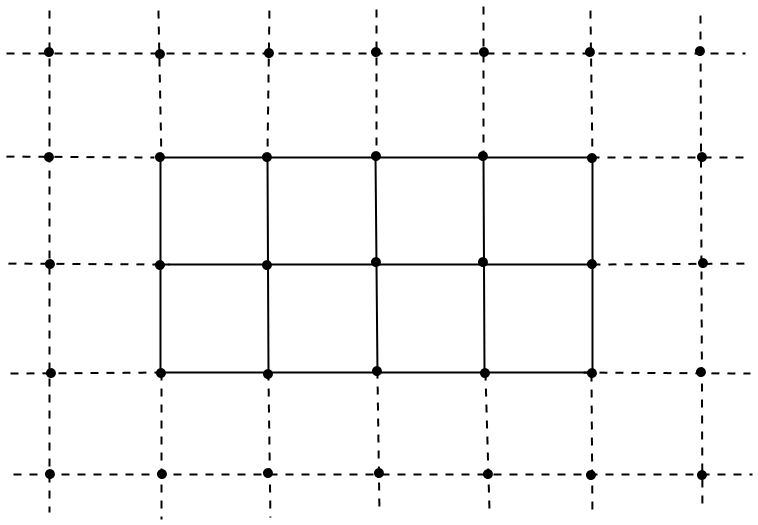}
    \caption{The subset of links represented by solid lines constitute the inside. The remaining links   shown by dashed lines are the outside.}
    \label{fig:}
  \end{center}
\end{figure}

Now consider a gauge-invariant state, $|\psi\rangle$, and some subset of all the spatial links. We will call these links as lying in the ``inside'', although our discussion is  general and does not require the collection of links to  form a closed or even connected region.  For a concrete example, consider Fig 1 with the subset of links of interest  lying inside and on the boundary of the rectangle.

As was discussed in the introduction there is an extended Hilbert space, ${\cal H}$, which is obtained in this case by taking the tensor product of all the two dimensional Hilbert spaces defined at each link. 
This Hilbert space by definition admits a tensor product decomposition in terms of the Hilbert spaces of links lying inside, ${\cal H}_{in}$, and outside (i.e. the rest of the links), ${\cal H}_{out}$. 
A basis for ${\cal H}$ is given by taking the  tensor product of the $|\pm 1\rangle$ basis introduced above for each link. Since the set of gauge-invariant states is a subset of ${\cal H}$, any state $|\psi\rangle$ is automatically embedded in ${\cal H}$. It is easy to see that the corresponding state in ${\cal H}$ is unique. For example, there is a unique way to expand $|\psi\rangle$ in terms of the basis obtained by taking the tensor product of the $|\pm 1\rangle$ states on each link. We will sometime refer to this as the natural embedding of $|\psi\rangle \in {\cal H}$ below. 
 
Our proposal is to take  the state $|\psi\rangle$, now regarded as an element of ${\cal H}$, and define the density matrix $\rho_{in}$ for the inside set of links by taking the trace,
\be
\label{defrhoin}
\rho_{in}=\Tr_{{\cal H}_{out}} |\psi\rangle\langle\psi|.
\ee
The entanglement entropy is then given by the von Neumann entropy associated with $\rho_{in}$, eq.(\ref{eein}). 
It is clear that this definition can be applied to any set of links.

Before proceeding let us also comment on some of the existing literature here. 
 A similar definition for the Abelian case was given in \cite{BP}. Also, in the context of an extended lattice
construction, as we will discuss further in section \ref{elc}, the same definition was given, both for the Abelian and Non-Abelian cases,  in\footnote{We thank W. Donnelly for correspondence in this regard.}  \cite{DD1}, \cite{D1}.


\subsection{Some Properties}
\label{properties}
Some comments are now in order about the properties  for the entanglement entropy which follow from our definition. 
Even though we state them in the $\mathbb{Z}_2$ context here, these properties are actually   generally true for all Abelian and Non-Abelian cases we study in this paper.

First, it is worth emphasising that the  definition is unambiguous.  Any gauge-invariant state corresponds to a unique state in ${\cal H}$,  as was mentioned above, and for any state in ${\cal H}$ there is a unique density matrix $\rho_{in}$, leading to a unique entanglement entropy.  In fact, one can similarly define the R\'{e}nyi entropies for any set of links and these too are all then unambiguously defined. 

Second, we note that the density matrix defined above $\rho_{in}$ allows one to calculate the expectation value of any gauge-invariant operator which only acts on the inside links. Any such operator, ${\hat O}$, can be promoted to be an operator acting on  ${\cal H}$.  And if it only acts on the inside links  then it is easy to see that 
\be
\label{expo}
\langle\psi|{\hat O}|\psi\rangle=\Tr(\rho_{in} O).
\ee
In this way we see that $\rho_{in}$ contains all the information needed  to  describe the results of any physical measurement   done on the inside links. 
Later on, in section \ref{morea}, we will see that the converse is also true, in the following sense.  Namely, we will argue that  measurements done on the inside links (more correctly an ensemble of inside link systems) can fully determine $\rho_{in}$. 

Third, the Hilbert space ${\cal H}$ obtained from the tensor product of the two-dimensional Hilbert spaces at each link also inherits from these $2$-dimensional subspaces a natural inner product, meeting all the standard properties of positivity etc. 
Thus the entanglement entropy we have defined automatically meets the strong subadditivity condition. If $A, B$ and $C$ are three sets of links which share no links in common then we have that 
\be
\label{ssa}
S_{A\cup B}+ S_{C\cup B} \ge S_B + S_{A \cup B \cup C}
\ee
where $S_{A \cup B}$ is the entanglement entropy for the set of links, $A\cup B$ etc. 

While these three properties have been discussed   in the context of the $\mathbb{Z}_2$ theory in this section, they will also continue to to be true for all the other theories considered in this paper essentially for the same reasons. 

\subsection{Equivalence With Other Definitions}
Here we show that the definition for entanglement entropy given above by embedding the state in the Hilbert space ${\cal H}$ agrees with the electric centre  choice of \cite{CHR} (see also \cite{Rad} for a clear discussion). 

To establish the equivalence we will need to re-express the entanglement entropy, eq.(\ref{eein}), as a sum over different, suitably defined, sectors. First, let us introduce some terminology. 
Vertices in the lattice will be said to  lie inside if all links ending on them are inside links, and outside if all terminating links are outside links. In addition there will be  boundary vertices which  have some terminating links lying  in the inside subset  and some lying  in the outside subset. 

To carry out the trace over ${\cal H}_{out}$ and define $\rho_{in}$ we can work in any basis. For  connecting to the electric centre prescription of CHR it is useful to consider the basis in which the   outside links ending on a   boundary vertex are in eigenstates of the electric flux operators defined on these links. For example, in Fig 2 we have shown the boundary vertex $V_1$ which has two inside links, labelled as $3,4$, and two outside links, labelled as $1,2$, ending on it. In the  basis we work with, the two outside links are in eigenstates of the $\sigma_1$ operators. This is of course not a complete specification of the basis of ${\cal H}_{out}$. But the remaining steps below  will not depend on specifying the basis for the other links in ${\cal H}_{out}$ explicitly, so we do not do so here.

\begin{figure}[t]
  \begin{center}
    \includegraphics[height=60 mm]{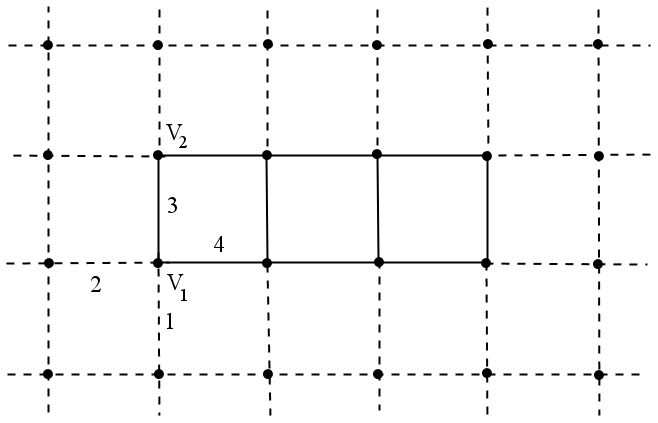}
    \caption{The lattice with boundary vertex $V_1$. Links 1 and 2 are in the outside subset, while   links 3 and 4 are in the inside subset. }
    \label{fig:2}
  \end{center}
\end{figure}
Now at any boundary vertex let us define the total electric flux operator going outside to be the product of all the $\sigma_1$ operators  acting on the outside links ending at this vertex. For example in Fig 2 vertex $V_1$ has two outside links, labelled $1,2$, with electric flux operators which we denote to be  $\sigma_{1}^{(1)}, \sigma_{1}^{(2)},$ respectively. The total electric flux operator  going outside at this vertex is then 
\be
\label{efva}
\sigma_{1}^{(1)}\sigma_{1}^{(2)}.
\ee
The basis we are working with therefore  consists of eigenstates of the total out-going electric flux operator at each vertex. We can then define a vector, ${\bf k}$, whose  entries,  $k_1,k_2, \cdots, $ give the eigenvalues of the total electric flux operators going outside at the boundary vertices $V_1, V_2, \cdots$, respectively.  This  vector specifies the electric flux which leaves the inside region at all the boundary vertices. 
We note that for the $\mathbb{Z}_2$ theory  each entry in ${\bf k}$ takes values $\pm 1$. 

It is  now clear that the Hilbert space of the outside links,  ${\cal H}_{out}$, can be written as  a direct sum over sectors of definite electric flux,
\be
\label{sumho}
{\cal H}_{out}= \bigoplus {\cal H}_{out}^{\bf{k}}.
\ee
By expressing the identity operator, ${\bf 1}$, of  ${\cal H}_{out}$, as a sum over projection operators ${\hat P}^{\bf{k}}$ into the subspace ${\cal H}_{out}^{\bf{k}}$,
\be
\label{idroj}
{\bf 1}= \sum_{\bf{k}} {\hat P}^{\bf{k}},
\ee
 we can express $\rho_{in}$ as
\be
\label{deffria}
\rho_{in}=\Tr_{{\cal H}_{out}} |\psi\rangle\langle\psi| =\sum_{\bf{k}} \Tr_{{\cal H}_{out}} \left( {\hat P}^{\bf{k}} |\psi\rangle\langle\psi|\right).
\ee
So that 
\be
\label{deffri}
\rho_{in}= \sum_{\bf{k}} \rho_{in}^{\bf{k}},
\ee
with, 
\be
\label{defrinka}
\rho_{in}^{\bf k}=\Tr_{{\cal H}_{ out}^{\bf k}} |\psi\rangle \langle\psi|,
\ee
where,
\be
\label{defrinkab}
\Tr_{{\cal H}_{ out}^{\bf k}} |\psi\rangle \langle\psi| \equiv  \Tr_{{\cal H}_{out}} \left( {\hat P}^{\bf{k}} |\psi\rangle\langle\psi|\right).
\ee

Any outside vertex by definition has only links lying in the outside subset ending on it. Therefore the operator ${\cal G}_{V_{i}}$, eq.(\ref{repgi}),  which generates gauge transformations at such a vertex, maps  ${\cal H}_{out}^{\bf{k}}$ to itself.
The subset of states in ${\cal H}_{out}^{\bf{k}}$ which are gauge-invariant with respect to all gauge transformations acting on the outside vertices will be denoted by ${\cal H}_{ginv, out}^{\bf{k}}$.
Since $|\psi\rangle$ is gauge-invariant, it is easy to see that only states lying in ${\cal H}_{ginv,out}^{\bf{k}}$ contribute to the trace in eq.(\ref{defrinka}).
This follows from the fact that  ${\cal G}_{V_{i}}$ has eigenvalues $\pm 1$,  and only states lying in ${\cal H}_{ginv,out}^{\bf{k}}$ have  eigenvalue $+1$ with respect to all outside gauge transformations. Thus we get
\be
\label{defrink}
\rho_{in}^{\bf k}=\Tr_{{\cal H}_{ginv, out}^{\bf k}} |\psi\rangle \langle\psi|.
\ee

The operators $\rho_{in}^{\bf{k}}$ act on ${\cal H}_{in}$. We can define the total electric flux carried inside at a boundary vertex, in an analogous fashion, to be the product of the electric flux operators for the inside links which end at this vertex. For example, for vertex $V_1$ which has inside links $3,4,$ ending on it the total electric flux operator is
\be
\label{definside}
\sigma_1^{(3)} \sigma_1^{(4)}.
\ee
And we can define a vector ${\bf{k}}$, now specifying the total electric flux carried in at each boundary vertex. 
Like ${\cal H}_{out}$, the inside Hilbert space, ${\cal H}_{in}$, can also be written as a direct sum, 
\be
\label{sumhin}
{\cal H}_{in}=\bigoplus {\cal H}_{in}^{\bf{k}}
\ee

Further, it is easy to see then that since the state $|\psi\rangle$ satisfies Gauss' law, eq.(\ref{gistrans}),  at  each boundary vertex, the total electric flux going out must equal the total electric flux going  in at each  boundary vertex. 
This means that the operator $\rho_{in}^{\bf{k}}$ will have non-zero matrix elements only for  states which lie in ${\cal H}_{in}^{\bf{\tilde k}}$ with 
${\bf k}={\bf {\tilde k}}$. 
As a result the entanglement entropy, eq.(\ref{eein}), takes the form, 
\be
\label{ee2}
S_{EE}=
- \sum_k \Tr_{{\cal H}_{in}^{\bf{k}} }\rho_{in}^{\bf{k} } \log(\rho_{in}^{\bf{k}}).
\ee
As a final step we note that  since $|\psi\rangle$ is invariant under gauge transformations acting on any inside vertex, only states  which are gauge-invariant with respect to such an inside gauge transformation can contribute in the trace on the RHS above. Denoting by ${\cal H}_{ginv, in}^{\bf{k}}$ the subset of all such gauge-invariant states in ${\cal H}_{in}^{\bf{k}}$, we get,
 \be
\label{fexee}
S_{EE}=-\sum_k \Tr_{{\cal H}_{ginv, in}^{\bf{k}} }\rho_{in}^{\bf{k} }\log(\rho_{in}^{\bf{k}}).
\ee

We are now ready to compare with the results of CHR. Their approach  deals with only gauge-invariant states and in particular with the algebra of gauge-invariant operators acting on these states. Their central point is that this algebra of gauge-invariant operators acting on  the inside links, ${\cal A}_V$,  has a non-trivial centre. Once the centre is diagonalised, each gauge-invariant subspace corresponding to a fixed set of eigenvalues for the centre 
does factorise,
  and the entanglement  entropy can be defined  by tracing over the outside Hilbert space. 
  
  Various choices can be made for  the centre. One choice, called the electric centre choice, corresponds to working with eigenstates of the total electric flux operator entering at each boundary vertex. We see that the different sectors, labelled by the vector, ${\bf{k}}$,  in terms of which we have expressed our result in eq.(\ref{fexee}) corresponds to exactly this choice. 
  In fact it is easy to see that our result in eq.(\ref{fexee}) exactly matches the result in eq.(27)  of CHR, see also eq.(3.12) of \cite{Rad},  for the entanglement entropy.  
  This follows after noting that we can define an  operator ${\tilde\rho}_{in}^{\bf{k}}$, of unit trace,  as,
  \be
\label{defpk}
\rho_{in}^{\bf{k}}=p_{\bf k} {\tilde \rho}_{in}^{\bf{k}},
\ee 
  with 
  \be
\label{tracea}
\Tr_{{\cal H}_{ginv, in}^{\bf{k}}}  \rho_{in}^{\bf{k}}= p_{\bf k}
\ee 
Eq.(\ref{fexee}) can then be rewritten in terms of ${\tilde \rho}_{in}^{\bf{k}}$ as, 
\be
\label{reee}
S_{EE}= -\sum_{\bf{k}} p_{\bf{k}} \log(p_{\bf{k}}) + \sum_{\bf{k}} p_{\bf{k}}  S(\tilde{\rho}_{in}^{\bf{k}})
\ee
where $S(\tilde{\rho}_{in}^{\bf{k}})$, the von Neumann entropy associated with ${\tilde \rho}_{in}^{\bf{k}}$, is 
\be
\label{vne}
S({\tilde \rho}_{in}^{\bf{k}})=-\Tr_{{\cal H}_{ginv, in}^{\bf{k}} }{\tilde \rho}_{in}^{\bf{k}} \log({\tilde \rho}_{in}^{\bf{k}}).
\ee
  
  Let us end this section with one comment. The different sectors, labelled by different values of the electric flux, ${\bf k}$, are actually different superselection sectors, as emphasised in CHR. Gauge-invariant operators acting on the inside links, or outside links,  alone, cannot change this electric flux and therefore  must preserve the sector to   which a state carrying this flux belongs.  We will return to this point in section \ref{more} when we discuss some measurement aspects of the  entanglement  entropy. 
  
  \section{ $U(1)$ Lattice Gauge Theory}
  \label{ua}
  Our discussion for the $\mathbb{Z}_2$ case readily extends to the $\mathbb{Z}_N$ and $U(1)$ cases. Here we will focus on the $U(1)$ case.  
  Once again we work in the Hamiltonian formulation and consider the theory on a spatial lattice which for definiteness we take to be a square lattice  in $2$ spatial dimensions.
 
Associated with  every link on the lattice in this theory is an angular degree of freedom. One difference is that the  links are  directed vectors in this case. 
The angular degree of freedom associated with the link $L_{ij}$ which starts from the vertex $i$ and ends in the vertex $j$ is $\theta_{ij}\in [0,2\pi]$. 
With the  oppositely oriented link $L_{ji}$  we associate the angle $\theta_{ji}=-\theta_{ij}$. 

Under an infinitesimal  gauge transformation at the vertex $V_i$,  parametrised by $\epsilon_i$,  the angular variables associated with all link variables emanating from $V_i$  are shifted by $\epsilon_i$,
\be
\label{gtua}
\theta_{ij}\rightarrow \theta_{ij}+\epsilon_i.
\ee

In the quantum theory 
associated with each link, $L_{ij}$, is a  Hilbert space, ${\cal H}_{ij}$. The angular variable is promoted to an operator ${\hat \theta}_{ij}$ acting on ${\cal H}_{ij}$.
 There is  also an operator ${\hat{\cal L}}_{ij}$ which  is  conjugate to ${\hat \theta}_{ij}$, 
\be
\label{defaa}
[{\hat{\cal L}}_{ij},{\hat\theta}_{ij}]=-i,
\ee
which can be thought of as the angular momentum associated with $\theta_{ij}$. It   generates shifts in the angle ${\hat \theta}_{ij}$:
\be
\label{shiftua}
e^{i \epsilon {\hat{\cal L}}_{ij}} {\hat \theta}_{ij} e^{-i \epsilon{\hat{\cal L}}_{ij} }={\hat \theta}_{ij} + \epsilon.
\ee
The eigenvalues of ${\hat{\cal L}}_{ij}$ are quantised since $\theta_{ij}$ is compact and takes values over the integers, $ [-\infty, .., -1, 0, 1, ..., \infty]$.
The operator ${\hat{\cal L}}_{ij}$  is  the electric flux associated with the $i-j$ link. 

An infinitesimal  gauge transformation  ${\cal G}_{V_{i}}(\epsilon)$, parametrised by $\epsilon$, is generated by the operator
\be
\label{defgtua}
{\cal G}_{V_i}= e^{i \epsilon \sum_j {\hat{\cal L}}_{ij}}
\ee
where the sum is over all links leaving the vertex $V_i$. 
Gauge-invariant states and operators  meet the conditions, eq.(\ref{gistrans}), and eq.(\ref{gino}).  


An underlying Hilbert space ${\cal H}$ is obtained by taking the tensor product of the Hilbert spaces ${\cal H}_{ij}$ defined at each link. The space of gauge-invariant states can be naturally embedded in ${\cal H}$, 
${\cal H}_{ginv} \subset {\cal H}$. 

For a pure state $|\psi\rangle$ and a collection of links which we call the inside ones, our proposal for the entanglement is completely analogous to the $\mathbb{Z}_2$ case. ${\cal H}_{out}, {\cal H}_{in}$, are the Hilbert spaces given by the tensor product of the outside and inside links. We regard $|\psi\rangle\in {\cal H}$, via the natural embedding discussed above, and define $\rho_{in}$ as in eq.(\ref{defrhoin}).
The entanglement entropy is then given by eq.(\ref{eein}).

We can show, just as in the $\mathbb{Z}_2$ case, that this entanglement can be written as a sum of terms each of which arises from a sector of definite outgoing  electric flux. This allows us to establish that our definition is the same as the one given in CHR and also \cite{Rad}. We adopt the terminology introduced above for the $\mathbb{Z}_2$ case, for inside, outside and boundary vertices. 
The one difference is that we  keep track of the orientation of the links, since they are directed vectors in the $U(1)$ case. We take the links which touch a boundary vertex to all be  oriented to emanate  from the vertex. 
To carry out the sum over ${\cal H}_{out}$ we then  go to a basis where the outside links  emanating from a boundary vertex are  eigenstates of the electric flux operators, ${\hat{\cal L}}_{ij}$. 
The total electric flux leaving a  boundary vertex and going out is then the sum of the eigenvalues of all the outside links emanating from this vertex. 

The Hilbert space ${\cal H}_{out}$ splits up into a direct sum of sectors, eq.(\ref{sumho}), each of which is specified by a vector, ${\bf k}$,  now with integer entries. The $ i^{\text{th}}$ entry is the total flux leaving the boundary vertex $V_i$. The inside density matrix can then be written as a sum, eq.(\ref{deffri}), with $\rho_{in}^{\bf{k}}$ being given by eq.(\ref{defrinka}). 
We can also define ${\cal H}^{\bf{k}}_{ginv,out}\subset {\cal H}^{\bf{k}}_{out}$ to be the subspace of states in flux sector ${\bf k}$ which are gauge-invariant with respect to all gauge transformations on the outside vertices. Then since only states belonging to ${\cal H}^{\bf{k}}_{ginv,out}$ contribute to the trace in eq.(\ref{defrinka}) we get eq.(\ref{defrink}). 

For the inside Hilbert space ${\cal H}_{in}$ we now specify the sectors by the vector ${\bf{k}}$ so that the $i^{\text{th}}$ entry  is {\it minus}  of the total electric flux carried by the inside links emanating from the $V_i$ boundary vertex.  It follows that ${\cal H}_{in}$ can also be written as the  direct sum, eq.(\ref{sumhin}). Since the state $|\psi\rangle$ satisfies Gauss' law at the boundary vertices, we learn that 
$\rho_{in}^{\bf{k}}$  has non-zero matrix elements with states in ${\cal H}_{in}^{\bf \tilde{k}}$ only when ${\bf{k}}={\bf\tilde{k}}$, leading to eq.(\ref{ee2}). And finally defining ${\cal H}_{ginv, in}^{\bf{k}}$ to be the subspace of states which are gauge-invariant with respect to gauge transformations on inside links, we can argue that the entanglement entropy is given by eq.(\ref{fexee}).

This result is in agreement with the definition given by CHR, see also \cite{Rad}. The different sectors which appear in the sum in eq.(\ref{fexee}) have different values for the electric flux, so we see
that the result agrees  with the  electric centre choice of CHR.

\section{The $SU(2)$ Gauge Theory}
\label{su2}
 The discussion above in turn can be generalised to  Non-Abelian gauge theories. As a concrete example we take  an $SU(2)$ gauge theory,  without matter, and again work in  the Hamiltonian framework, in $2$ spatial dimensions.  The degrees of freedom then  correspond to  a  $2\times 2$ matrix $U_{ij} (\vec{\theta})\in SU(2)$ associated with each link $L_{ij}$ (more abstractly a group element of $SU(2)$).  We see that the matrix is parametrised by three angles $\vec{\theta}$ which can be thought of as the  configuration space of a rigid rotor, see \cite{KS}. To avoid confusion, let us emphasise that the $i,j,$ indices in $U_{ij}$ refer to the link $L_{ij}$ of the spatial lattice with which it is associated, and are not the matrix indices of the $2\times 2$ matrix which we have suppressed in our condensed notation. The matrix $U^\dagger_{ij}= U_{ji}$ will be associated with the oppositely oriented link denoted by $L_{ji}$. 
 
 Under an infinitesimal gauge transformation  at the vertex $V_i$ the matrix variables corresponding to all links emanating from $V_i$  transform as 
 \be
 \label{gtsu}
 U_{ij}\rightarrow \left(1+ i \epsilon^a {\sigma^a \over 2}\right)  U_{ij},
 \ee
 where $\epsilon^a=1,2,3$ are infinitesimal parameters, and $\sigma^a$ are the Pauli matrices.
 
 In the quantum theory there is a Hilbert space ${\cal H}_{ij}$ of states associated with each link. The matrix variables $U_{ij}$ are promoted to operators, ${\hat U}_{ij}$, which act on ${\cal H}_{ij}$. 
 Eigenstates of ${\hat{U}_{ij}}$ will be denoted by $|U_{ij}\rangle$ and provide a basis of ${\cal H}_{ij}$. 
 We have 
 \be
 \label{evsu}
 {\hat U}_{ij} |U_{ij}\rangle=U_{ij} |U_{ij}\rangle
 \ee
 where $U_{ij}$ is the corresponding matrix in $SU(2)$. 

 There are also operators $\hat{J}^a_{ij}, a=1,2,3,$ which generate $SU(2)$ transformations:
 \be
 \label{gen}
 e^{i \epsilon^a \hat{J}^a_{ij}} {\hat U}_{ij} e^{-i \epsilon^b \hat{J}^b_{ij}}= \left(1- i  \epsilon^a {\sigma^a\over 2}\right) \hat{U}_{ij},
 \ee
 where $\epsilon^a$ are infinitesimal parameters. 
  On the RHS above these matrices act on the left. Acting on the eigenstates, $|U_{ij}\rangle$ we get,
  \be
  \label{actesl}
  e^{i \epsilon^a \hat{J}^a_{ij}} |U_{ij}\rangle= \left\lvert\left(1+ i \epsilon^a {\sigma^a \over 2}\right) U_{ij}\right\rangle
  \ee
 Similarly one can define operators, denoted by ${\hat {\cal J}}^a_{ij}$,   which  result in an infinitesimal group element acting on the right,
 \be
 \label{actesr}
 e^{i \epsilon^a {\hat {\cal J}}^a_{ij}}|U_{ij}\rangle= \left|U_{ij}\left(1- i \epsilon^a {\sigma^a\over 2}\right)\right\rangle
 \ee
 (note the minus sign on the RHS).

 The operators $\hat{J}^a_{ij}$, for each link, $L_{ij}$,  satisfy the standard angular momentum algebra,
 \be
 \label{angmom}
 [\hat{J}^a_{ij}, \hat{J}^b_{ij}]= i \epsilon^{abc} \hat{J}^c_{ij}.
 \ee
 Similarly, from eq.(\ref{actesr}) we find that the operators ${\cal {\hat J}}^a$ satisfy the algebra,
 \be
 \label{anm2}
 [{\hat {\cal J}}^a, {\hat {\cal J}}^b]=i \epsilon^{abc} {\hat {\cal J}}^c,
 \ee
  see, e..g.,  \cite{GY}.
  Also, 
 from eq.(\ref{actesl}), eq.(\ref{actesr}) it also follows that ${\hat J}^a, {\cal {\hat J}}^b, a,b, =1,2,3$ commute, 
 \be
 \label{comm}
 [{\hat J}^a, {\cal {\hat J}}^b]=0.
 \ee

 
 Next, consider the vertex $V_i$ and orient all the links which touch this vertex to be emanating from it. Then a gauge transformation at this vertex is generated by the operator
 \be
 \label{gtsua}
 {\cal G}_{V_i}= e^{ i \epsilon^a \sum_j J^a_{ij}}
 \ee
 where the sum is over all the $4$  links which emanate from $V_i$. 
 Gauge-invariant states  $|\psi\rangle$ and  operators ${\cal O}$ will then meet the conditions, eq.(\ref{gistrans}) and eq.(\ref{gino}). 
 
 It is worth mentioning that the angular momentum operators ${\hat J}^a_{ij}$ are not the direct analogues of the electric flux operators here. Instead the electric flux carried by the $L_{ij}$ link is measured by the difference,
 \be
 \label{diff}
 {\hat Q}^a_{ij}= {\hat J}^a_{ij}-{\hat {\cal J}}^a_{ij},
 \ee
 see \cite{KS}. This difference arises because   gluon degrees carry charge in the Non-Abelian case. 
 
 Consider the directed link $L_{ij}$ which emanates from the vertex $V_i$ and ends on $V_j$. The matrix valued degree of freedom $U_{ij}$ which lives on this link is acted on by both the operators ${\hat J}^a$ and ${\hat {\cal J}}^a$, which commute with each other, eq.(\ref{comm}).  The operator ${\hat J}^a$ enters in the gauge transformation ${\cal G}_{V_i}$, eq.(\ref{gtsua}). Similarly the operator ${\cal {\hat J}}^a$ will enter in the gauge transformation at the vertex $V_j$. This follows after noting that   
 ${\hat J}^a$ and ${\cal {\hat J}}^a$  generate the left  and right actions on $U_{ij}$, eq.(\ref{actesl}),  eq.(\ref{actesl}) respectively.

 Let us also mention in passing that we can define operators $({\hat U}_{(1, ij)})^{ab}, a,b =1,2,3$, which acting on the eigenstates $|U_{ij}\rangle$ of the Hilbert space $H_{ij}$ on link $L_{ij}$ give
 \be
 \label{actua}
 ({\hat U_{(1, ij)} })^{ab}|U_{ij}\rangle= U^{ab}_{(1, ij)} |U_{ij}\rangle
 \ee
 where $U^{ab}_{(1, ij)}$ is the matrix element of the spin $1$ (3-dimensional) representation of $SU(2)$ corresponding to group element $U_{ij}$. 
 One can then show that
 \be
 \label{rellr}
 {\hat {\cal J}}^a = -{\hat J}^b {\hat U}_{(1)}^{ba}.
 \ee
 The state on the link $L_{ij}$ can then be specified by the quantum numbers, $ J^2=j(j+1),  J^3, {\cal J}^3$, for example, since these operators commute with each other. 
 Note that the choice of the $J^3, {\cal J}^3$ components is purely a matter of convention here; we could have equally well chosen other components instead.

 The extended Hilbert space we work with for the entanglement entropy is the tensor product of the Hilbert spaces on each of the links, ${\cal H} = \bigotimes {\cal H}_{ij}$,
 Similarly ${\cal H}_{out}$ and ${\cal H}_{in}$ are the Hilbert spaces obtained by taking the tensor product of the outside and inside link Hilbert space. 
 As before, for this case too, we obtain a density matrix, $\rho_{in}$ by taking the natural embedding of  the state $|\psi\rangle\in {\cal H}$ and then tracing over ${\cal H}_{out}$, eq.(\ref{defrhoin});
 and define its entanglement entropy to be eq.(\ref{eein}). 
 
 It is interesting to try and express our result  as a sum over suitably defined  sectors in this case as well. 
  As before we will use the term  boundary vertex to denote a vertex that has some inside links and some outside links touching it. 
 At each boundary vertex we orient all links to be emanating from it. 
 And define the total outgoing angular momentum, $\hat{J}^a_{out,T}$,  to be the sum of the angular momenta of all the outside links at this vertex. E.g., at vertex $V_1$ shown in Fig 1. Let the two outside links be denoted as $1,2$, with angular momentum $\hat{J}^a_1, \hat{J}^a_2$ respectively, then 
 \be
\label{defjout}
\hat{J}^a_{out, T}\equiv \hat{J}^a_1+\hat{J}^a_2. 
\ee
 The states of the outside links can be organised into representations of definite outgoing angular momentum, $(\hat{J}_{out,T})^2$.
 And the Hilbert space, ${\cal H}_{out}$, can be written as a sum over  sectors of definite values for $(J_{out,T})^2$.  This can be done at each boundary vertex, which in general could have $1,2$ or $3$, outside links emanating from it, resulting in a  total specification of the  different sectors  by giving  the quantum number at each boundary vertex for
 $(J_{out,T})^2$, where $\hat{J}^a_{out,T}$ 
 at each vertex is the total angular momentum carried by the outside links.   The resulting sectors are therefore specified again by a column vector, ${\bf k}$, with the $i^{\text{th}}$ entry specifying the value of $(J_{out,T})^2$ at the boundary vertex, $V_i$.
 

 In a similar way ${\cal H}_{in}$ can also be expressed as a direct sum over sectors. 
We define the total  ingoing  angular momentum, ${\hat J}^a_{in,T}$ at a  vertex to be the total angular momentum carried by the ingoing links which emanate from a boundary vertex. 
E.g., for $V_1$, with the ingoing links labelled as $3,4$, we have, 
\be
\label{defjin}
\hat{J}^a_3 + \hat{J}^a_4=\hat{J}^a_{in,T}. 
\ee
Then  the Hilbert space ${\cal H}_{in}$ can be written as a sum over  sectors of definite $(J_{in, T})^2$. And Gauss' law tells us that for the state $|\psi\rangle$
\begin{eqnarray}
(J_{out,T})^2& = & (J_{in,T})^2 \label{glsua}\\
 \hat{J}_{out,T}^a |\psi\rangle& = & -\hat{J}_{in, T}^a |\psi\rangle, a= 1,2,3. \label{glsub}
 \end{eqnarray}
 We can therefore specify the different sectors which appear in ${\cal H}_{in}$ also  by a vector $\bf{k}$ whose entries now 
 stand for the values of $(J_{in,T})^2$ at the boundary vertices.
 
 It is worth noting that the difference between the operators ${\hat J}^a, {\cal {\hat J}}^a$, is important in ensuring that ${\cal H}_{in}$  can be expressed as a direct sum over different sectors \footnote{In general this is true for ${\cal H}_{out}$ also, although not for the specific set of inside links considered in Figure 2.}. For example,  consider the link labelled as ``$3$'' in  Figure 2, oriented to emanate from vertex $V_1$ and end on vertex $V_2$.  The operator ${\hat J}^a_3$ will then enter in the definition of the ingoing angular momentum at vertex $V_1$, as discussed above, whereas the operator ${\cal {\hat J}}^a_3$ will enter in the definition of the total ingoing angular momentum for vertex $V_2$. These commute, eq.(\ref{comm}), thereby ensuring that ${\cal H}_{in}$ can be expressed as a direct sum over sectors. More generally, consider the link $L_{ij}$ which goes from $V_i$ to $V_j$ and the oppositely oriented link $L_{ji}$. With orientation $L_{ij}$ we have operators ${\hat J}^a_{ij}, {\cal {\hat J}}^a_{ij}$, etc.
 Then, it follows from the definitions, eq.(\ref{actesl}), and (\ref{actesr}), that 
 \be
 \label{reltwoj}
 {\hat J}^a_{ij}={\cal {\hat J}}^a_{ji}, \ \ a=1,2,3.
 \ee

 Before proceeding, let us note that  a more complete description  of the different sectors of ${\cal H}_{in}$ is possible, by specifying both $(J_{in,T})^2$, and  in addition one component, say $J^3_{in,T}$, at every boundary vertex.  
 However, as we will discuss below, this is not necessary because $\rho_{in}^{\bf{k}}$ will end up commuting with ${\hat J}^a_{in,T}$, and thus will have no non-trivial dependence on this extra quantum number. This also makes it superfluous to keep track of additional subsectors in ${\cal H}_{out}$ which arise for a fixed value of $(J_{out,T})^2$, and  different values of  $J^3_{out,T}$.

 The Hilbert spaces, ${\cal H}_{out}, {\cal H}_{in}$,  can then be written as a direct sum over different sectors of total outgoing and incoming angular momentum at each boundary vertex, eq.(\ref{sumho}), eq.(\ref{sumhin}).
 The rest of the arguments are essentially identical to the earlier cases. The trace over ${\cal H}_{out}$ splits up into different subspaces ${\cal H}_{out}^{\bf{k}}$ and in each subspace only the gauge-invariant states (i.e. states which are gauge-invariant with respect to gauge transformations acting on outside vertices) contribute. For a sector specified by the vector $\bf{k}_1$, say,
 the resulting operator, $\rho_{in}^{\bf{k_1}}$, when acting on ${\cal H}_{in}$  only has non-zero matrix elements in the subspace ${\cal H}_{in}^{\bf{k_1}}$, due to the gauge-invariance of the state $|\psi\rangle$ we started with. Moreover, the operator does not have any nonzero matrix elements with states which are not gauge-invariant with respect to gauge transformations defined on the inside vertices. From these observations it then follows that the resulting entanglement entropy can be written in the form given in eq.(\ref{ee2}) and eq.(\ref{fexee}).

To return to a comment made above, it is easy to see, from Gauss' law, eq.(\ref{glsub}), and the invariance of the trace under a change of basis, that $\rho^{\bf{k}}_{in}$ satisfies the relation,
\be
\label{relaax}
[{\hat J} ^a_{in,T}, \rho^{\bf{k}}_{in}]=0,
\ee
 at every boundary vertex, and for all $\bf{k}$. 
Thus $\rho^{\bf{k}}_{in}$ behaves like a singlet operator  under ${\hat J} ^a_{in,T}$, and has no nontrivial dependence on the $J^a_{in,T}$ quantum numbers within a sector of fixed $\bf{k}$.

\subsection{Discussion}
\label{dissu2}
 We see that our definition  yields an expression for the entanglement entropy which is similar in form to that obtained in the Abelian cases discussed earlier. 
 
 A few differences are worth taking note of though.  
 The different sectors over which the sum must be carried out do not correspond to specifying the total electric flux carried away at the boundary vertices, by the outside links, as was mentioned above, since the operators $\hat{ J}_{out,T}^a$ do not  directly measure this flux. 
 Also, we argued above that we do not need to  keep  track of different subsectors specified by the additional quantum number,  $J^3_{in,T}$,  say, in each sector with fixed $(J_{in,T})^2 $, as far as  
 $\rho^{\bf k }_{in}$
 is concerned,  since eq.(\ref{relaax}) is met. However, these subsectors do get important when we discuss  more operational ways to measure entanglement and compare them with our definition,
 see section \ref{more}. In fact each subsector, corresponding to a different value of $J^3_{in,T}$, is a  different superselection sector by itself , which cannot be changed by acting with gauge-invariant  operators  on the inside links. This is   because any such operator must commute with $\hat{J}_{out,T}^a$, and Gauss' law, eq.(\ref{glsub}), relates $\hat{J}_{out,T}^a$ to  $\hat{J}_{in,T}^a$.

  It is also worth commenting on our  result for the Non-Abelian case vis-a-vis  the approach of CHR, \cite{CHR}, which has been discussed above.  
  For the Abelian case the entropy obtained with  our definition agrees with that obtained in   the electric  centre choice  definition  of CHR. The different sectors, specified by different values for  $\bf{k},$ which arise in the sum in eq.(\ref{fexee}), correspond to different  values for the electric flux operators, which are a choice for the centre of the algebra of gauge-invariant operators, ${\cal A}_V$. And in this way  our result directly maps to that obtained in \cite{CHR} using the electric centre choice.

  
  
  In the Non-Abelian case things are similar, as we have seen above. Eigenvalues of the operator, $({\hat J}_{out,T})^2$, or equivalently $({\hat J}_{in,T})^2$, label the different sectors in the expression for the entanglement entropy.  And this operator is gauge-invariant, although it does not directly correspond to the electric flux. 
   In this way our result can be related  to a suitable choice of  a centre in the algebra of gauge-invariant operators, ${\cal A}_V$, as suggested by CHR.  
  
  For specifying the different superselection sectors, mentioned above, one can no longer work with just gauge-invariant operators. Instead one also needs to specify say the  $J^3_{in,T}$ quantum numbers, which are not gauge-invariant. 
  In the  extended Hilbert space, ${\cal H}_{in}$,  which includes  non-gauge-invariant states, one can define a bigger algebra containing also non-gauge-invariant operators. And the operator $\hat{J}^3_{in,T}$, along with 
  $(\hat{J}_{in,T})^2$ (which is gauge-invariant)  can be taken to lie  in the centre of such a bigger algebra. In this way,  the discussion of superselection sectors  could also be perhaps   related to the general approach of CHR, but now extended to apply to  operators in the bigger Hilbert spaces  ${\cal H}$, ${\cal H}_{in}$.

 
\section{Replica Trick}
\label{replica}
 We will now show that our formula for the entanglement entropy agrees with a definition of the replica trick. 
  Connections to the replica trick are also discussed in \cite{BP}.
 
 The replica trick is based on carrying out a Euclidean path integral in space-time. For the $(2+1)$-dimensional theories we have been studying so far in the Hamiltonian formulation, the path integral is over $3$-dimensional Euclidean manifolds. To define the R\'{e}nyi entropy $R_n= \frac{1}{1-n} \log \Tr\rho_{in}^n$ one calculates a  path integral over a manifold where the degrees of freedom are not single-valued in $\mathbb{R}^3$  but on its $n$-fold cover. This $n$-fold cover  is obtained by introducing a branch cut  at a particular instant of ``time,'' along the boundary of the  spatial region of interest, and identifying the values for the variables below the cut in one sheet with the values they take above the cut in the adjoining next sheet. A definition for the entanglement entropy is then given  by suitably continuing the result for the R\'{e}nyi entropy for $n \rightarrow 1$. 
 
 We can apply this procedure readily to the lattice gauge theories we have been studying in this paper. The  gauge theories  have been analysed in the Hamiltonian formulation,  living on a spatial lattice  with continuously varying time. The path integrals  which  arises for such theories most naturally, in the context of our present discussion,  are  also  in the continuum   in time and on a  lattice in the space directions. The degrees of freedom in these path integrals are link variables, living on the spatial links, and  varying along the time direction.  E.g., in the $U(1)$ case the link variable is an angle, $\theta_{ij}(t)\in [0,2\pi]$, as explained in section \ref{ua}, while  in  the $SU(2)$ case it is a $2\times 2$ matrix $U_{ij}(t) \in SU(2)$, see section \ref{su2}. 
 
 Now
consider a set of links, e.g., the set of of all the links inside and on the boundary of the rectangle shown in figure 1.  To calculate the entanglement entropy  of this set of links with the rest, one introduces a cut, say at $t=0$, 
 so that in the path integral the values these link variables  take  just above the cut, at $t=0^{+}$, are different from their values just below the cut, at $t=0^{-}$. 
 Then by calculating the path integral on the $n$-fold cover, and taking the appropriate  limit,  as mentioned above,  one obtains the entanglement entropy.  
 
 We will now  argue that the result obtained in this way for a gauge theory in fact agrees with our definition of the entanglement entropy given in the previous sections.  
 
 \subsection{General Background} 
 To see this,  let us first go back to the general case and understand why the path integral on the $n$-fold cover  is expected to  give the R\'{e}nyi entropy $R_n$. We consider a situation where the cut is introduced
 at $t=0$, as in the gauge theory case above, along the boundary of the spatial region of interest. 
 The important point is that path integral from $(-\infty, 0^{-})$  essentially prepares an appropriate state at $t=0^{-}$, which we denote by $|\psi\rangle$.  Similarly the path integral from $(0^{+}, \infty)$ gives  rise to $\langle\psi|$. 
 This follows  by  relating the path integral to the evolution induced in Euclidean time  due to a Hamiltonian, in the standard fashion. With standard boundary conditions, the lowest energy state dominates,
 resulting in $|\psi\rangle$ being the vacuum, $|0\rangle$, of the system.

 For example, take   a scalar field theory (which could be on a spatial lattice to make the analogy with the gauge theory case more direct). And  let us choose a basis of eigenstates of the field operator $\hat{\phi}({\bf x},t)$  which we denote as   $|\phi({ \bf x},t)\rangle$. Then  we get that 
 \be
 \label{inint}
 \langle\phi({\bf x},t=0^{-})|\psi\rangle=\int_{t=-\infty}^{\phi({\bf x},0^{-})} [D \phi] e^{-S}
 \ee
 where $S$ is the action and we are not being very explicit about the initial conditions at $t\rightarrow -\infty$. A similar expression relates $\langle\psi|\phi({\bf x}, 0^{+})\rangle$ to a path integral from $t=0^+$ to $\infty$,
 \be
 \label{outing}
 \langle\psi|\phi({\bf x},t=0^+)\rangle=\int_{\phi({\bf x},0^{+})}^{t=\infty} [D \phi] e^{-S}.
 \ee
 The full path integral  involves integrating over $\phi({\bf x}, t)$ while  taking  the values  of the field variables  at $t=0$ to be continuous in the outside region, i.e. 
 \be
 \label{condphia}
 \phi({\bf x},0^-)=\phi({\bf x},0^+)=\phi({ \bf x},0), {\bf x}\in {\rm outside},
 \ee
and taking the values of $\phi({\bf x},t) $ in the outside region to be discontinuous, taking values, $\phi({\bf x}, 0^-)$ and $\phi({\bf x}, 0^+)$ below and above the cut.
   It is then clear from eq.(\ref{inint}) and eq.(\ref{outing}) that this gives rise to the  matrix element,
 \be
 \label{matrixele}
 \langle\phi( {\bf x}, t=0^{-}) |\rho | \phi( {\bf x},t=0^{+})\rangle= \int_{t=-\infty}^{t=+\infty } [D \phi] e^{-S},
 \ee
 with ${\bf x}\in {\rm inside}$. In eq.(\ref{matrixele})
  $\rho$ is the density matrix of the inside region, and the path integral is carried out in the presence of the cut. 
 
Once  this identification between the path integral on one sheet and the matrix element of the density matrix is established, it follows immediately that the R\'{e}nyi entropy $R_n$ is given by the path integral over the $n$-fold cover. And the entanglement entropy can then be calculated from the formula, 
 \be
 \label{eerep}
 S_{replica}=-[(n \partial_n-1)\log \Tr\rho^n] |_{n=1}.
 \ee
assuming the continuation to $n\rightarrow 1$ is well defined. 
A minor point is that the path integral does not usually give rise to a normalised state $|\psi\rangle$ of unit norm, but eq.(\ref{eerep}) takes this into account to give the  von Neumann entropy of the  correctly normalised density matrix with unit trace, see, e.g., \cite{S1}.

 \subsection{Gauge Theories}
 We can now return to the case of gauge theories. For this case a basis which is the analogue of $|\phi({\bf x},t)\rangle$ above can be introduced as follows. We start with  eigenstates of the operator $\hat{U}_{ij}$ at each link. 
 The operator $\hat{U}_{ij}$  for the $SU(2)$ theory was introduced in section \ref{su2}   and  its eigenstates were denoted by $|U_{ij}\rangle$, with  the eigenvalues $U_{ij}$ being elements of $SU(2)$.  Similarly, an operator $\hat{U}_{ij}$ can be defined in the $U(1)$ case with its eigenvalues  $U_{ij}$ being elements of $U(1)$. 
 A full basis  can  then be obtained by taking the  tensor product of all such basis elements on each link. An element of this basis is given by $\bigotimes_{(ij)}|U_{ij}\rangle$.
 This basis is the analogue of $|\phi({\bf x},t)\rangle$. 
  A basis for the inside links is obtained by taking the 
  tensor product only over the basis elements for  the inside links, which we schematically write as $\bigotimes_{(ij)\in in} |U_{ij}\rangle$. Similarly a basis for the outside is given by 
  $\bigotimes_{(ij) \in out}|U_{ij}\rangle$. 
 
 It is then easy to see that the  discussion  above applies also for these gauge theories. The path integral from $(-\infty, 0^-)$, and $(\infty, 0^+)$,  gives rise to the  state $|\psi\rangle$ and $\langle\psi|$ respectively.
 With eq.(\ref{inint}) taking the form, 
 \be
 \label{psigtc}
 \langle U_{ij}(0^{-})|\psi\rangle= \int_{t=-\infty}^{U_{ij}(0^{-})} [D U_{ij} ] \ \  e^{-S}
 \ee
 and similarly for eq.(\ref{outing}). 
 For the $SU(2)$ case the functional integral for $U_{ij}$ on the RHS  of eq.(\ref{psigtc}) involves the standard Haar measure on $SU(2)$. 
 
 The full path integral is obtained by also integrating over  the outside link variables at $t=0$, we denote these by $U_{ij}(t=0)$ with $(ij)\in out$ .
 These variables are continuous across $t=0$. This gives rise to the density matrix $\rho$ acting on the inside links. As a result,  the path integral with $U_{(ij) \in inside}$ taking values, 
 $U_{(ij)\in inside}(0^-)$, at $t=0^-$ and 
 $U_{(ij)\in inside}(0^+)$, at $t=0^+$ respectively, 
 leads to the matrix element
  \be
 \label{mela}
 \langle U_{(ij)\in in}(0^{+})|\rho|U_{(ij)\in in}(0^{-})\rangle=\int [D U_{ij}] e^{-S}.
 \ee
It then follows  that the $n$-fold cover  computes the R\'{e}nyi entropy $\Tr\rho^n$ and thus this definition of the  replica trick can be interpreted as the von Neumann entropy for  $\rho$. The action $S$ is related to the Hamiltonian
in the standard way by replacing the momenta by time derivatives of $U_{ij}$. 
 
 Note that the path integral on the RHS preserves gauge-invariance with respect to spatial gauge transformations, i.e. gauge transformations where the parameters vary only in space and not time. This follows from the invariance of $S$ and the invariance of the Haar measure under these gauge transformations. If the initial state at $t=-\infty$ is also gauge-invariant, or in effect gauge-invariant as would happen if the vacuum dominates,  then $|\psi\rangle$ at $t=0^-$ would also be gauge-invariant and 
 one can show from eq.(\ref{psigtc})  that 
 \be
 \label{ginvpath}
 \langle U_{ij}|\psi\rangle=\langle U^g_{ij}|\psi\rangle
 \ee
 where $U_{ij}, U^g_{ij}$ are related by the gauge transformation $g$. 
 
 Usually the  path integral formulation for lattice gauge theories is presented in a somewhat different manner,  corresponding to the Euclidean formulation, with both time and the spatial coordinates being on a lattice. Starting from this formulation one can obtain the above type of path integral representation by taking the continuum limit along the $t$ direction and fixing $A_0=0$ gauge. These steps are identical to those taken in obtaining the Hamiltonian formulation. The resulting theory then only has invariance with respect to spatial gauge transformations. Note that no further gauge fixing is needed to make the path integral in eq.(\ref{mela}) well defined, since the integrals involve the Haar measure on  compact groups and are finite.
 
 We are now ready to show that  the replica trick definition given above agrees  with  our definition of the entanglement given in the previous sections. In fact to do so we simply have to note that the discussion above has, in effect,  been carried out with $|\psi\rangle$ being  embedded  in the  extended Hilbert space, ${\cal H}$. This extended Hilbert space, we remind the reader,   includes  in particular non-gauge-invariant states. This  is because the basis $\bigotimes_{(ij)}|U_{ij}\rangle$ is in fact a basis of ${\cal H}$ and similarly $\bigotimes_{(ij)\in in} |U_{ij}\rangle$ and 
 $\bigotimes_{(ij)\in out} |U_{ij}\rangle$ are a basis for ${\cal H}_{in}$ and ${\cal H}_{out}$ respectively. 
 Thus the density matrix defined by the path integral  is actually $\rho_{in}=\Tr_{{\cal H}_{out} }(|\psi\rangle\langle\psi|)$, eq.(\ref{defrhoin}), and the R\'{e}nyi entropies and von Neumann entropy obtained from the replica trick agree with those  obtained by taking the trace of suitable powers of $\rho_{in}$ in ${\cal H}_{in}$. This shows that the answer one gets from the replica trick  defined above agrees  with our definition in  eq.(\ref{eein}) for the entanglement entropy.
 
 We have considered the $SU(2)$ and $U(1)$ cases above in $3$ space-time dimensions, for definiteness,   but it is clear  that the   argument applies  more generally, in higher dimensions,  and also for other Non-Abelian gauge groups, and Abelian discrete groups.  
 
  Let us end this section with one comment. On the lattice   other definitions of the replica trick are possible, different from the one we have used above. One expects that in the continuum limit all these definitions agree unto counter terms, including some localised on the boundary of the region of interest.\footnote{We thank Gautam Mandal and Shiraz Minwalla for a discussion on this point.}

\section{More on The Entanglement }
\label{more}
\subsection{Measuring the Entanglement }
\label{morea}

It is well known in the literature of quantum computation that the entanglement in a bipartite system can be quantified operationally  by comparing  it to the entanglement of a reference system,  usually taken to be  a set of $N$ Bell  pairs.\footnote{ We thank John Preskill for insightful  comments which led  us to examine issues discussed in this subsection.} This comparison can be done in two ways, see, e.g., \cite{NC}, \cite{JP}. 
For example, in the process of entanglement distillation, one  starts with the entangled parts $A,B$, of a bipartite pure system and an extra set of unentangled qubits. The two subsystems $A,B,$ are assumed to be separated in space. We then  ask, in an asymptotic sense with a large number of copies of the initial system, how many pairs of entangled Bell pairs can be produced by transferring the entanglement between $A,B$ to the qubits, using only local operations and classical communication. For a system with local degrees of freedom, e.g.,  a spin system, it is known that this number $N$ is given by 
\be
\label{odenta}
 N=-\Tr(\rho \log \rho)
\ee
where $\rho$ is the density matrix of $A$ or of $B$. 
Alternatively, in entanglement dilution, we take the two parts $A,B$ to be nearby,   and a reservoir of Bell pairs such that  one partner in each pair is  near the system $AB$, while the other is far away.   We then ask how many Bell pairs would be needed for teleporting one of the two parts, say A,  to the distant location. The answer,  for the number of such Bell pairs, $N$, in the asymptotic sense mentioned above,  for a system with local degrees of freedom is again given by eq.(\ref{odenta}). 

We would now like to examine whether  for  gauge theories under consideration here the result of such a comparison with Bell pairs  also agrees with our definition of the entanglement entropy given in eq.(\ref{eein}). In fact, this  turns out  not  to be true. The entanglement entropy which we have defined   is not equal to the number of Bell pairs involved in entanglement distillation, or  dilution, in general.  The two parts $A,B$  of the  system, are the   set of inside and outside links, as per the definitions used above. The  essential obstacle is that the local operations which one can use  correspond to  gauge-invariant observables  acting on the inside or the outside links alone. We saw above that our result for the entanglement entropy can be expressed as a sum over different superselection sectors. Gauge-invariant local operators cannot connect these different sectors, and acting with only these operators is then not enough  to make  the full entropy available for a comparison with  Bell pairs.\footnote{In gauge theories  it is still true that for a  large number of copies, $n$,  of the system of interest one can   work in effect in a subspace of reduced dimensionality, $2^p$, where $p=n S_{EE}$,
and $S_{EE}$ is the full entanglement entropy given by eq.(\ref{eein}). This smaller dimension subspace also admits a tensor product decomposition between $A$ and $B$.  But the process of distillation requires a further unitary transformation which transfers the entanglement between $A, B,$ in this reduced subspace to a set of Bell pairs. For gauge theories this unitary transformation in general cannot be carried out using gauge-invariant operators alone, since these operators  cannot connect different superselection sectors. As a result,  the full entanglement entropy we have defined, eq.(\ref{eein}), is not available for distillation. }

In the Abelian case the different superselection sectors correspond to different values of electric flux, and this observation about the full entropy not being available for the comparison was also made by CHR. 
Working within a flux sector labelled by ${\bf k}$ the entropy can be converted into entangled Bell pairs, or alternatively used in teleportation. But the  rest of the information  in $S_{EE}$,
eq.(\ref{reee}),  having to do with the probabilities $p_{\bf{k}}$ is not directly related to the standard measures of entanglement.  Note that a dependence on $p_{\bf{k}}$ enters both in the first term and also the second term in eq.(\ref{reee}) through the coefficients, $p_{\bf{k}}$, so neither of these terms or their sum is directly the result of such a  comparison. 

In the Non-Abelian case, as was mentioned in section \ref{su2}, the different superselection sectors do not correspond just to different values of ${\bf k}$, and therefore to different values of $(J_{in,T})^2$.
 Rather, the different superselection sectors are given by subsectors within each  sector of fixed ${\bf k}$. These subsectors  correspond to different values of a component of the in- coming total angular momentum,   say $J^3_{in,T}$,  for a fixed value of $(J_{in,T})^2$, at each boundary vertex. It would be worth understanding this difference between the Abelian and Non-Abelian cases
 further.  

 It  is  also worth  mentioning that even though our   answer for the entanglement, eq.(\ref{eein}),  does not directly correspond to the number of Bell pairs which are obtained  in entanglement distillation or dilution,  the density matrix $\rho_{in}$ we obtain can be  completely measured by local experiments performed  on the inside links alone, as long as  one  has  an ensemble of many  copies of the inside system available to us. For example, in the Abelian case, one can measure the electric flux on each of the links, and also transitions induced between different values of these electric fluxes by inserting  Wilson loops lying in the inside, and  these are enough to determine $\rho_{in}$ completely. 
 
 The situation for the Non-Abelian case is similar. We saw in eq.(\ref{relaax})  that $\rho_{in}^{\bf{k}}$ is a singlet under the total incoming angular momentum, 
 ${\hat J}^a_{in,T}$, at each boundary vertex. For determining $\rho_{in}^{\bf{k}}$ in each sector,  from which the  full density matrix can be obtained, it is therefore enough to obtain the values of $\Tr(\rho_{in}^{\bf{k}} {\hat O})$ for a complete set of Hermitian operators which are singlets under ${\hat J}^a_{in,T}$. In fact  gauge-invariant observables acting on the inside links provide such a complete set, and therefore allow us to determine $\rho_{in}$ fully.

 To see why gauge-invariant operators acting on the inside are singlets we note that 
 a gauge-invariant operator acting on the inside must commute with the total outgoing angular momentum at each boundary vertex, ${\hat J}^a_{out,T}$, by definition because the operators only act on the inside links. And they must also commute with the total angular momentum, ${\hat J}^a_{T}= {\hat J}^a_{in,T}+ {\hat J}^a_{out,T}$,
 since these are the generators of gauge transformations. Thus they must commute with ${\hat J}^a_{in,T}$, which means they are singlets with respect to these operators. A little more thought shows that they are also a complete set of such singlet operators.

\subsection{Entanglement For Spatial Regions}
In the introduction, we  began  by considering a region of space and asking about its entanglement, but then  chose to focus on the  entanglement for a set of links instead. Now that we have understood the situation for a collection of links, to some extent, it is worth going back and asking about the entanglement for a region of space. Suppose for concreteness that  we are interested in the  region of space enclosed by the curve in Fig 3, which has one connected component.
 There are links leaving this region, which intersect the curve; these are among the  dashed links  shown in Fig 3. To use our definition  for the entanglement entropy in such a situation  we need to decide whether to include the intersecting links   in  the set of interest (the inside as per our definition above).  

To be specific let us take the case   of an Abelian gauge theory. 
We will see below that for some links which intersect this curve the answer for the entanglement  of the inside set of links does not change whether or not they are included. 
 But for other links it can  change. 
 
 \begin{figure}[t]
   \begin{center}
     \includegraphics[height=85mm]{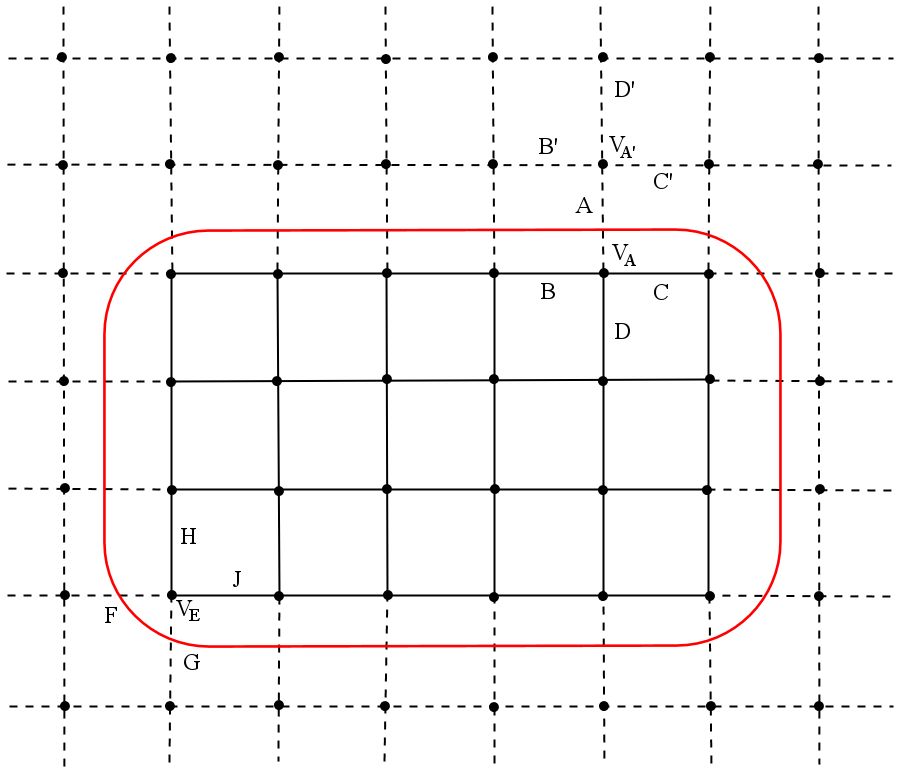}
     \caption{The red curve is the boundary of the spatial region of interest.  Several links intersect it, including link A which emanates from vertex $V_A$ and  and links F,G, which emanate from vertex $V_E$.}
     \label{fig:3}
   \end{center}
 \end{figure}
First, consider link A,  as shown in Fig 3. This is the only link which connects  the boundary vertex $V_A$ to the outside. 
Gauss' law  at vertex $V_{A}$ fixes the value of the electric flux at this  link in terms of the electric fluxes carried by links $B,C,D$ which are inside. Similarly Gauss' law on vertex $V_{A'}$ which is outside and the other end-point of link A fixes the value of the  flux on link A in terms of the flux at the three other  links, B',C',D',  which are outside and also end on $V_{A'}$. 
This means that the electric flux on A is completely correlated with degrees of freedom inside {\it and}  outside. It then follows simply that including this link variable in the inside or the outside does not change the entanglement entropy eq.(\ref{eein}). 
Next, consider the vertex $V_{E}$ which is connected to the outside by two links F,G, and to the other inside vertices by H,J. It is now clear that Gauss' law at $V_{E}$ only relates the total electric flux   on F,G, with that  on  H,J, and does not fix the individual values of the two electric fluxes  on F,G. 
This means there are degrees of freedom on the F,G, links which  are   not fully correlated with  the inside and outside (in the $\mathbb{Z}_2$ case this is one qubit's worth).  As a result,  the entanglement we obtain by including or not including links F, G, will  be  different in general. 

 The  example   above shows that  ambiguities arise when we attempt  to   extend our definition of the entanglement entropy  to apply to   a  spatial region  instead of a collection of links.

 \begin{figure}[t]
   \begin{center}
     \includegraphics[height=75 mm]{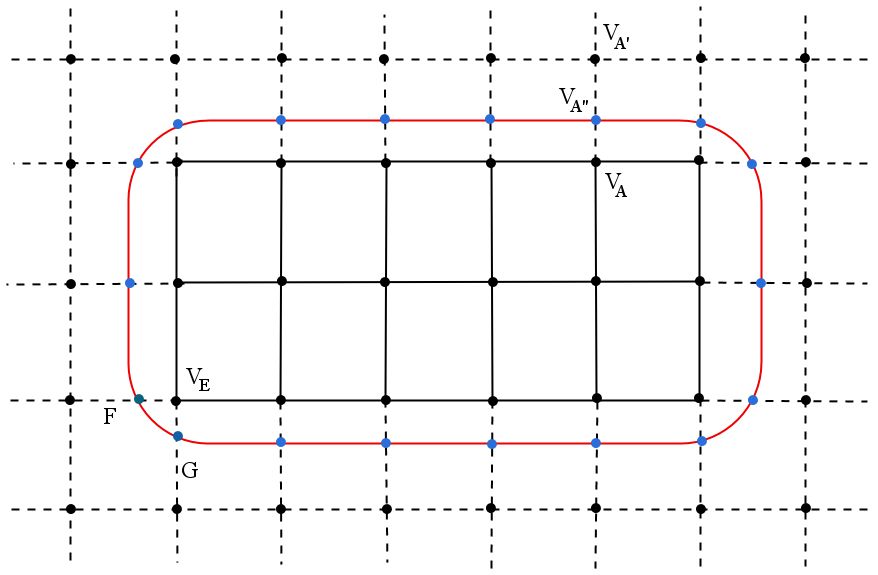}
     \caption{The extended lattice construction. Blue dots represent additional  vertices  which have been  introduced at the points where  links of the original lattice intersect the  boundary  shown in red. Links  touching the blue dots are the new links which have been   introduced.}
          \label{fig:4}
   \end{center}
 \end{figure}
 
 \subsection{The Extended Lattice Construction}
 \label{elc}
 
 Another definition for the entanglement entropy both in Abelian and Non-Abelian  gauge theories has been
 proposed in the literature, see, \cite{HIZ},  \cite{DD1}, \cite{BP}, \cite{D1}, \cite{D3}. 
 Following \cite{CHR}, which has a helpful summary that we will use here,  we will refer to this as the extended lattice construction (ELC).  
 This definition applies to a region of space whose boundary does not go through any vertex of the lattice. In figure 4 we show such a region which lies  in the interior of the boundary red curve.
 The boundary  is intersected by links which connect the inside to the outside. The idea behind the definition is to extend the lattice by introducing extra links and vertices
 such that the enlarged  Hilbert space associated with this extended lattice then admits a tensor product decomposition. This is done by introducing extra vertices  at points where links intersect the boundary, and splitting each link which intersects the boundary into two by allowing it to terminate at the extra vertex that has been introduced. 
 For example, in figure 4  consider the link which connects $V_A$ and $V_{A'}$, for brevity we  refer to it as the $A-A'$ link below. We introduce a new vertex $V_{A''}$ where this link intersect the boundary and then split the $A-A'$ link into two, giving rise to the $A-A''$, and $A''-A$ links. The extra vertices introduced in this way are  shown in blue in figure 4. 
 
 In the ELC definition the Hilbert space associated with the extended lattice, ${\cal H}'$,  consists of states which satisfy the Gauss' law constraint at the original vertices, shown in black in Fig 4, but which need not satisfy this constraint on the extra vertices which have been introduced, shown in blue. As a result it turns out that  ${\cal H}'$ now admits a tensor product decomposition. 
 
 To understand this better, let us introduce the terminology of ``inside links'' to refer to links which either lie securely  inside the boundary  or to  links which connect vertices lying inside with the extra vertices which lie on the boundary. Outside links will then refer to all remaining links. For example, in figure 4 the inside links  include the link $A-A''$, but not the link $A''-A'$ which lies in the outside. 
 It is then easy to see that since the states in ${\cal H}'$ do not necessarily meet the Gauss' law constraint on the extra vertices we have the tensor product decomposition 
 \be
 \label{tneelc}
 {\cal H}'={\cal H}_{in}' \otimes {\cal H}_{out}'.
 \ee
 Here ${\cal H}_{in}'$ is the Hilbert space of states defined on inside links which are gauge-invariant with respect to inside vertices but not the extra vertices lying on the boundary curve.
 And similarly  ${\cal H}_{out}'$ is the Hilbert space of states on outside links which are gauge-invariant with respect to the outside vertices but not the vertices on the boundary curve. 
 
 A  state $|\psi\rangle$ which lies in ${\cal H}_{ginv}$ - the Hilbert space of gauge-invariant states defined on the original lattice - is then embedded as a state $|\psi'\rangle \in {\cal H}'$ in a well defined way, given by eq.(\ref{psipelc}).
 Although ${\cal H}'$  includes in general states which are not gauge-invariant with respect to gauge transformations at the extra vertices, as mentioned above,  all states, $|\psi'\rangle$ which arise in ${\cal H}'$  due to embedding in this manner are actually gauge-invariant.  
 As an example consider the links $A-A''$, $A''-A'$, in Fig 4, and let the link variables on them take values $U_{A-A"}, U_{A"-A'}$, respectively. 
     Then the  state $|\psi'\rangle\in {\cal H}'$  is defined by its wave function satisfying the relation    
     \be
     \label{psipelc}
     \psi'(U_{A-A''}, U_{A''-A'})=\psi(U_{A-A''}\cdot U_{A''-A'}),
     \ee
     where ``$\cdot$'' denotes matrix product. Here
     we are suppressing the dependence of  $\psi$  on the other link variables. 
     Under a gauge transformation parametrised by group element $g$  at the extra vertex $A''$, 
     \be
     \label{gacta}
     U_{A-A''}\rightarrow U_{A-A''} \cdot g^{-1}, \ \  U_{A''-A'}\rightarrow g \cdot U_{A''-A'}. 
     \ee
     As a result we see that $|\psi'\rangle$ is gauge-invariant.\footnote{This discussion applies also to the
 Non-Abelian case, see, \cite{D1},  for which the order in which $g$ acts in eq.(\ref{gacta}) is important.} One minor comment is that $|\psi'\rangle$ as defined in eq.(\ref{psipelc}) does not necessarily have  unit norm.  This can be taken care of by introducing a normalisation constant on the RHS of eq.(\ref{psipelc}), or in the case of continuous groups,
like the $U(1)$ or $SU(N)$ examples,   by defining the measure for summing over the group  to be of unit norm (as is the case with the Haar measure). 

 
 Since ${\cal H}'$ admits a tensor product decomposition one then defines the entanglement entropy of the inside links  by tracing over ${\cal H}_{out}'$ to obtain a density matrix, 
 \be
 \label{defrhoelc}
 \rho=Tr_{{\cal H}_{out}'}|\psi'\rangle\langle\psi'|,  
 \ee
 and then equating the von Neumann entropy of $\rho$,
 \be
 \label{vnelc}
 S=-Tr_{{\cal H}_{in}'} \rho \log(\rho),
 \ee
  with the entanglement entropy.

 We will now argue that both for Abelian and Non Abelian cases the ELC definition above agrees with our definition, eq.(\ref{eein}) when applied to the extended lattice. We have also argued that in the Abelian case
 our definition is equivalent to the electric centre choice of CHR. It will then follow that for the Abelian case the ELC definition is equivalent to the electric centre choice as well,  when applied to the extended lattice, as was noted in \cite{CHR} already. 
 
 The equivalence with our definition is in fact straightforward. 
    Consider applying the definition we have proposed, eq.(\ref{eein}), to  the extended lattice described above, with the choice  of the  inside  and outside subsets of links,  for purposes of applying our definition, being the same as what was given 
   above in the ELC case.  And with the gauge-invariant state with  which our analysis begins  being the state $|\psi'\rangle$ on the extended lattice.\footnote{This means that for applying the equations in  section \ref{z2}, e.g., eq.(\ref{deffria}), we replace $|\psi\rangle$ by $|\psi'\rangle$.} It is then easy to see that  in fact ${\cal H}_{out}'$, ${\cal H}_{in}'$,  introduced above can be expressed as the direct sums, 
  \be
 {\cal H}_{out}' =  \bigoplus_{\bf{k}} {\cal H}_{ginv,out}^{\bf k} \label{elta},
 \ee
 and,
 \be
  {\cal H}_{in}'  =  \bigoplus_{\bf{k}}{\cal H}_{ginv,in}^{\bf{k}} \label{eltb},
  \ee
  where the subspaces ${\cal H}_{ginv,out}^{\bf{k}}$, ${\cal H}_{ginv,in}^{\bf{k}}$  were introduced in section \ref{z2}. 
  Since $\rho_{in}$  is  given by  eq.(\ref{deffri}), with $\rho^{\bf k}$ being defined in eq.(\ref{defrink}), we see  from eq. (\ref{elta}) that $\rho_{in}$ is in fact the same as $\rho$ given in eq.(\ref{defrhoelc}). Similarly the trace in eq.(\ref{fexee}) can be seen to be the same as the trace over ${\cal H}'_{in}$, thereby leading to the entanglement entropy being the same.

  We end with one comment. The ELC definition pertains to the extended lattice and does not directly yield the answer for the entanglement  entropy  of  a set of links in the original unextended lattice. For example,  consider the set of links shown in bold face which lie inside the red curve  in figure 4.
    It is easy to see that the entanglement entropy  which  arises using our definition for this set of links  is different  from the entanglement entropy for the extended lattice associated to it, in general. This difference arises in the Abelian case itself, and is also present in the non Abelian case.


  Let us begin with the case of an Abelian theory, and  let us consider the situation for the $A-A'$ link in the original lattice. As argued in the previous subsection, in the original lattice, the electric flux carried by link $A-A'$ is correlated completely with the total electric flux on the inside links which terminate at $V_{A}$ and  also the outside links terminating at 
 $V_{A'}$. In the extended lattice, given the  mapping proposed  in the ELC construction for a  state in the extended Hilbert space , one can show that the value of the  electric flux carried by the $A-A"$ and $A"-A$ links  are the same, and equal to the value of the electric flux on the  $A-A'$ link   in the original lattice. It therefore follows that the two new links, $A-A"$ and $A"-A'$, of the extended lattice also take values completely 
correlated with the total electric flux   carried by the inside links at the $V_{A}$ vertex and  the other outside links at the $V_{A"}$ vertex respectively. 
It is then easy to see that adding these two links, $A-A"$, $A"-A'$, to the inside and outside sets does not change the entanglement. 
 However, if we instead consider vertex $V_E$, also discussed in the previous subsection, the conclusion is different. Now in the original lattice only the total electric flux on the outside links $F,G$ is 
fixed by Gauss' law applied to the vertex $V_{E}$. The individual fluxes on the links $F, G$ are not fixed. As a result, one finds  that when both of these link are split into two, and the procedure in the ELC is carried out, 
the entanglement we get will  in general be different from the one calculated by using our definition (or equivalently the electric centre choice of CHR) on the original lattice. 
 
In the non-Abelian case also there is a difference between the ELC definition and the one we propose. In fact the difference is if anything even more pronounced and arises even for a link of the  
$A-A'$ typed discussed above, due to the fact that more quantum numbers are needed to specify the state of a link completely in this case.  In the $SU(2)$  gauge theory  for example, the  state  of the
$A-A'$  link is specified by $3$ quantum numbers, $(J)^2, J^3, {\cal J}^3$, as discussed in section \ref{su2}. 
 Gauss' law at $V_{A}$ fixes $(J)^2, J^3$, and at vertex $V_{A'}$, $(J)^2=({\cal J})^2, {\cal J}^3$, in terms of the total angular momentum of the inside links and the other outside links  emanating from $V_A, V_{A'}$ respectively. 
 When the link is divided into two, as per the definition of the ELC construction,  the $(J)^2, J^3$ numbers of the $A-A"$ link are fixed by the total   angular momentum of the inside links at $V_{A}$ and the 
$({\cal J})^2, {\cal J}^3$ quantum numbers 
of the $A"-A'$ link are  fixed by the total angular momentum of the other outside links at the vertex $V_{A'}$. However Gauss' law  does not  automatically determine the ${\cal J}^3$ quantum number of $A-A"$, and the $J^3$ quantum number of $A"-A'$ in terms of the angular momenta of the inside links on $V_{A}$ and the outside links on $V_{A'}$ respectively. Gauss' law at the new vertex $A"$  does tells us that these two additional quantum numbers must take equal values though and  they are therefore fully correlated. It is then easy to see that including $A-A"$   in the inside and  $A"-A'$  in the outside will in general gives a result which is different from what is obtained using our definition for the original  set of links on the unchanged lattice.

  \section{Conclusion}
  \label{conc}
  In this paper we have proposed a definition for the entanglement entropy of a gauge theory. Our definition applies to both Abelian and Non-Abelian gauge theories on a spatial lattice. 
  In the Abelian case without matter, we show that the definition is equivalent to the electric centre choice of CHR, \cite{CHR}. In the Non-Abelian case,  we find that the situation is similar and the  definition is tied to diagonalising operators lying in the  centre of   the algebra of gauge-invariant  operators, ${\cal A}_V$,  defined on the links of interest. We also show that in general, both for the Abelian and Non-Abelian cases, our definition agrees with the result obtained by using  a definition of the replica trick. This is a good argument for taking the definition given here seriously.
  
    
  Our definition applies to any subset of links on the spatial lattice and results in  standard desirable  properties, like strong subadditivity. 
  Interestingly, we find that the definition we propose, does {\it not}  agree with standard ways of quantifying entanglement discussed in the quantum information theory literature. These ways are based on using a collection of Bell pairs as a standard measure of entanglement. An entanglement entropy is assigned between two parts of a system (the inside and outside links) by comparing them with a  collection of Bell pairs, using the process of entanglement distillation or dilution which involve  local operations and classical communication. 
  The  difference between our definition  and such a  more operational definition has to do with the presence of superselection sectors. 
  These superselection sectors cannot be changed by local gauge-invariant operations acting on  only one of the two parts, and this prevents the whole entanglement from being made available  for the process of comparison. For Abelian theories the different superselection sectors correspond to different values of the electric flux, as noted by CHR. 
  In the Non-Abelian case  this is not true any more. In fact  the situation is even more interesting since  the superselection sectors are  labelled  by quantum numbers some of which are not gauge-invariant.
   This non-gauge-invariance drops out of the final answer  for the entanglement which is obtained by summing over all sectors.  
  
  We find this  mismatch between  our definition and the more operational ways to quantify entanglement  quite interesting. At root, the mismatch  arises because  gauge theories have non-local gauge-invariant operators, like electric and magnetic flux loops, and correspondingly non-local excitations. It might therefore be a general feature of attempts  to define the entanglement entropy in gauge theories. 
  
  There are many other ways in which the work reported here can be extended. This includes analysing the continuum limit, other measures of entanglement, like mutual information and relative entropy,
  and  a study of  gauge theories with matter. There is a good argument now showing that the Ryu-Takayanagi (RT) proposal in AdS gravity theories, \cite{RT1}, \cite{RT2},  follows from the replica trick in the  dual field theory, \cite{ML}. The definition we give, by embedding the state in an enlarged Hilbert space, ${\cal H}$,  can be applied to matter theories as well. 
  And it seems quite reasonable that the equivalence between our definition and the replica trick would continue to hold in these cases too. In this way one can  hope to firmly establish the equivalence between our definition and the RT proposal in AdS gravity. 
  
  
  Let us end by mentioning some  more topics for further study. It would be  worth thinking about the entanglement entropy for a spatial region, as opposed to a set of links, further, since preliminary attempts on the lattice at least lead to ambiguities, as was  discussed above. Also, the behaviour of the  entanglement entropy under a  duality transformation, which exchanges a gauge theory with a spin system in $3$ dimensions, for example, is  worth elucidating further. Finally,  the connection to the  extended lattice construction, as discussed 
in \ref{elc}, too bears further thought.
  
 \section{Acknowledgements}
 We acknowledge discussions with Shamit Kachru, Karthik  Inbasekar, Nilay Kundu,  Gautam Mandal, Shiraz  Minwalla, Rickmoy Samanta,  Nilanjan Sircar, Suvrat Raju, Anirudh Krishna, Pranjal Nayak, K Narayan and Debangshu Mukherjee. We  especially thank \DJ or\dj e Radi\v cevi\'c for helpful discussions and   John Preskill and   Spenta Wadia for their insightful comments.  
 SPT acknowledges support from the CERN  Theory Division for a sabbatical visit from July-December 2014 where  this work was done. 
He also acknowledges support from the J. C. Bose fellowship of the Government of India and  from the Department of Atomic Energy, Government of India. 
Most of all, we thank the people of India for their generous support.

 










\begin{thebibliography}{99}

\bibitem{CHR} 
  H.~Casini, M.~Huerta and J.~A.~Rosabal,
  Phys.\ Rev.\ D {\bf 89}, 085012 (2014)
  [arXiv:1312.1183 [hep-th]].
  
\bibitem{Rad} 
  \DJ.~Radi\v cevi\'c,
  arXiv:1404.1391 [hep-th].
  
  \bibitem{Srednicki} 
  M.~Srednicki,
  Phys.\ Rev.\ Lett.\  {\bf 71}, 666 (1993)
  [hep-th/9303048].

\bibitem{HLW} 
  C.~Holzhey, F.~Larsen and F.~Wilczek,
  Nucl.\ Phys.\ B {\bf 424}, 443 (1994)
  [hep-th/9403108].
  
  \bibitem{DFM} 
  L.~De Nardo, D.~V.~Fursaev and G.~Miele,
  Class.\ Quant.\ Grav.\  {\bf 14}, 1059 (1997)
  [hep-th/9610011].

\bibitem{CC1} 
  P.~Calabrese and J.~L.~Cardy,
  J.\ Stat.\ Mech.\  {\bf 0406}, P06002 (2004)
  [hep-th/0405152].
  
  \bibitem{CC2} 
  P.~Calabrese and J.~Cardy,
  J.\ Phys.\ A {\bf 42}, 504005 (2009)
  [arXiv:0905.4013 [cond-mat.stat-mech]].
  
  \bibitem{CH1} 
  H.~Casini and M.~Huerta,
  J.\ Phys.\ A {\bf 42}, 504007 (2009)
  [arXiv:0905.2562 [hep-th]].
  
  \bibitem{S1} 
  S.~N.~Solodukhin,
  Living Rev.\ Rel.\  {\bf 14}, 8 (2011)
  [arXiv:1104.3712 [hep-th]].
  
\bibitem{Kabat} 
  D.~N.~Kabat,
  Nucl.\ Phys.\ B {\bf 453}, 281 (1995)
  [hep-th/9503016].
  
  \bibitem{HIZ}
  A. Hamma, R. Ionicioiu and P. Zanardi, 
  Phys. Rev. A {\bf 71}, 022315 (2005)
  [quant-ph/0409073].
  
  \bibitem{DD1}
  W. ~Donnelly, 
  Phys. Rev. D {\bf 77}, 104006
(2008) 
  [arXiv:0802.0880 [gr-qc]]
 
 \bibitem{BP}
 P. ~V. ~Buividovich and M. ~I.~Polikarpov, 
  Phys. \ Lett. \ B {\bf 670}, 141 (2008)
[arXiv:0806.3376 [hep-th]] 

  \bibitem{D1} 
  W.~Donnelly,
  Phys.\ Rev.\ D {\bf 85}, 085004 (2012)
  [arXiv:1109.0036 [hep-th]].
  
\bibitem{D2} 
  W.~Donnelly and A.~C.~Wall,
  Phys.\ Rev.\ D {\bf 86}, 064042 (2012)
  [arXiv:1206.5831 [hep-th]].
  
  
  
  \bibitem{S2} 
  S.~N.~Solodukhin,
  JHEP {\bf 1212}, 036 (2012)
  [arXiv:1209.2677 [hep-th]].
  
  
\bibitem{D3} 
  W.~Donnelly,
  Class.\ Quant.\ Grav.\  {\bf 31}, no. 21, 214003 (2014)
  [arXiv:1406.7304 [hep-th]].
  
\bibitem{KS} 
  J.~B.~Kogut and L.~Susskind,
  Phys.\ Rev.\ D {\bf 11}, 395 (1975).
  
\bibitem{Kogut} 
  J.~B.~Kogut,
  Rev.\ Mod.\ Phys.\  {\bf 51}, 659 (1979).
  
  \bibitem{GY}
  K. Gottfried and T-M Yan, ``Quantum Mechanics: Fundamentals'',  Springer Verlag, 2nd edition, section 7.5(d).
  
   \bibitem{NC}
   M. A. Nielsen and I. L. Chuang, \emph{Quantum Computation and Quantum Information}, Cambridge University Press, Cambridge, 2000.
  
  \bibitem{JP}
  John Preskill, ``Lecture Notes On Quantum Computation,''
  \url{http://www.theory.caltech.edu/~preskill/ph229}
  

  
  \bibitem{RT1} 
  S.~Ryu and T.~Takayanagi,
  Phys.\ Rev.\ Lett.\  {\bf 96}, 181602 (2006)
  [hep-th/0603001].

  
\bibitem{RT2} 
  T.~Nishioka, S.~Ryu and T.~Takayanagi,
  J.\ Phys.\ A {\bf 42}, 504008 (2009)
  [arXiv:0905.0932 [hep-th]].
  
  
  
   
  \bibitem{ML} 
  A.~Lewkowycz and J.~Maldacena,
  JHEP {\bf 1308}, 090 (2013)
  [arXiv:1304.4926 [hep-th]].
  
    
  




\end{thebibliography}
\end{document}